\documentclass{article}

\usepackage{hyperref}
\usepackage{bookmark}
\usepackage{multicol}
\usepackage{graphicx}
\usepackage{amsfonts}
\usepackage{amsmath}
\usepackage{natbib}

\begin{document}
\title{Viral content propagation in Online Social Networks} 

\author{
  Giannis Haralabopoulos\\
  University of Southampton\\
  \texttt{Giannis.Haralabopoulos@soton.ac.uk}
  \and
    Ioannis Anagnostopoulos\\
  University of Thessaly\\
  \texttt{janag@dib.uth.gr}
  \and
   Sherali Zeadally\\
  University of Kentucky\\
  \texttt{szeadally@uky.edu}
  }
 
 \maketitle

\begin{abstract}
Information flows are the result of a constant exchange in Online Social Networks (OSNs). OSN users create and share varying types of information in real-time throughout a day. Virality is introduced as a term to describe information that reaches a wide audience within a small time-frame. As a case, we measure propagation of information submitted in Reddit, identify different patterns and present a multi OSN diffusion analysis on Twitter, Facebook, and 2 hosting domains for images and multimedia, ImgUr and YouTube. Our results indicate that positive content is the most shared and presents the highest virality probability, and the overall virality probability of user created information is low. Finally, we underline the problems of limited access in OSN data.
\end{abstract}

\section{Problem Definition}

Online Social Networks (OSNs) are modern information systems that allow users to interact with friends, acquaintances, family members, colleagues or even unknown individuals. The information exchanged in these networks is diverse and includes local or world news, general and scientific facts, quotes, personal preferences, and so on. OSN providers focus on developing new information-sharing schemes where users can easily disseminate information efficiently.

Information propagation and diffusion fields emphasize the temporal aspects of information systems. The scale and usability of OSNs have introduced new terms to describe the propagation phenomenon. Rapid sharing of information on a large scale is informally known as virality. The dynamics of information describe the properties that constitute, sustain, or modify the topology of the OSNs based on diffusion and propagation processes such as hierarchy, network partition, clustering, and others. User influence characterizes the capacity of OSN nodes to be a compelling force on behaviour and opinion shaping.

Our study focuses on content originating from Reddit, mainly because this OSN is comprised of both original and reused content, and it offers a liberal data access policy, and an open Application Programming Interface (API) with extensive documentation. Reddit is a social news and entertainment site powered by user generated content. Registered users submit content through a descriptive link, subject to (up/down)votes. Users who have created a post or commented on one, gain or lose "karma", a Reddit metric for user ranking. This metric is calculated as, the sum of all the upvotes minus the sum of all the downvotes a user receives. Posts that acquire a high vote ratio (positive to negative) in a short time period after their submission, are moved to the front page. It is apparent that Reddit community defines the popularity of the disseminated content and determines its "success" or "failure".

Information is shared and spread among various OSNs, such as Twitter, Facebook, and Google Plus. Information submitted in any of these OSNs affects the others as well. Our goal is to measure this impact, along with its properties and tempo-spatial attributes. We present the two main patterns observed in information propagation, based on whether the content existed before or was created for the particular OSN submission. A wider range of sub-patterns is presented in \cite{haralabopoulos2015lifespan}.

\section{Related Work}

The topic of diffusion has been at the center of sociology interest for many years. Even before the emergence of OSNs, social ties and information flows have been studied in traditional real-life social networks.

The notion that information flows, from mass media to opinion leaders and later on to a wider population as final consumers, was first introduced during the middle of 40's \cite{lazarsfeld1948peoples}. Almost a decade later, Kaltz and Lazarsfeld revisited the subject by proposing their theory of "Two step flow of communication" \cite{katz1966personal}. In a similar conceptualization, Granovetter and Mark suggested the analysis of social networks, as a tool for linking micro and macro levels of sociology theory \cite{granovetter1973strength}. Similar conclusions were verified in OSNs after nearly 40 years, while at the same time, interest in social media analysis skyrocketed \cite{bakshy2012role,rajyalakshmi2012topic}.

Viral marketing was introduced in 1997 by Juvertson and Draper \cite{jurvetson1997viral}.
Porter and Golan in \cite{porter2006subservient}, found that provocative content is a crucial virality factor. Leskovec et al. in \cite{leskovec2007cost}, modelled outbreak detection via node selection, while performing a two-fold evaluation of their model. Concerning word of mouth scenarios, Allsop et al. in \cite{allsop2007word}, noted that 59\% of individuals frequently share online content. Berger and Milkman in \cite{berger2012makes}, observed that positive content is more viral than negative.

The authors in \cite{gomez2010inferring} focused on identifying the optimal network that best describes information propagation in news media and blogs. In \cite{iribarren2011affinity}, authors analysed word of mouth through email. Social influence modeling was studied in \cite{goyal2010learning}. Similarly, the authors in \cite{ienco2010meme}, investigated the social influence in meme social graph. In \cite{yang2010predicting}, the authors studied virality lifespan issues in Twitter, as also mentioned in \cite{ienco2010meme}. In the same OSN (Twitter), Sakaki et al. \cite{sakaki2010earthquake} studied users as social sensors by monitoring information flow and dissemination during earthquake incidents. In \cite{kwak2010twitter}, the authors analysed the entire Twitter graph in order to assess its topological characteristics. 

In \cite{tucker2014reach}, Tucker measured the effectiveness of virality as a marketing tool. While in \cite{jurvetson1997viral} direct messages were measured. Interactions had a negative effect on virality while sharing, engaging, provocative (as defined in \cite{porter2006subservient}) or humorous content and visual appeal, affected virality in a positive way. In \cite{nahon2011blogs}, the authors created a map of the so-called "life cycle" of the blog. Likewise, a virality study in OSNs conducted by Guerini et al., showed that content is more important to virality than the influencer itself \cite{guerini2011exploring}. By focusing on diffusion analysis of Twitter, Hansen et al. in \cite{hansen2011good}, observed the effect of negativity on virality. In other OSN-related researches, the authors of \cite{bonchi2011influence} studied the maximization of influence. The work described in \cite{romero2011influence} examines the passivity of Twitter users. In \cite{mathioudakis2011sparsification}, the authors developed an algorithm to describe influence propagation.

Rajyalakshmi et al. in \cite{rajyalakshmi2012topic}, proposed a stochastic model for the diffusion of several topics. The authors discovered that strong ties play a significant role in virality, having homophily as a major contributor as observed in \cite{kwak2010twitter}. Micro and macro scales of the network become relevant, as seen in \cite{granovetter1973strength}. Furthermore, the authors of \cite{kwak2010twitter} note that, acts within groups have a global impact. Romero et al. in \cite{romero2011differences}, analyse hashtag diffusion in Twitter and discovered that initial adopters of hashtags were fairly important. Yang and Leskovec also studied content exposure growth and how it fades over time in Twitter \cite{yang2011patterns}. Huang et al. in \cite{li2013selection} tried to effectively select a number of nodes in order to monitor information diffusion. In a different but of related context study, Guerini et al. in \cite{guerini2012linguistic} noted that the dynamics of Social Networks exist in the scientific literature and used psycholinguistic analysis to determine how abstracts affect virality. 

The authors of \cite{guille2012predictive}, studied the dynamics of information diffusion process. Furthermore in \cite{bakshy2012role}, Bakshy et al. addressed the problem of information diffusion in OSNs. The authors ended up with the statement stated in \cite{granovetter1973strength,kwak2010twitter} that, strong ties are more influential but weak ties are responsible for propagating novel information. In \cite{karnik2013diffusion}, the authors investigated why and how some messages in OSNs become viral. Finally, Weng et al. in \cite{weng2013role}, analysed longitudinal microblogging data.

In this research, we aim to quantify the linkage of various OSNs based on contextual analysis, as well as to measure the diffusion in these OSNs, thus addressing a gap in the related literature. Based on the related works discussed, in this work, we:
\begin{itemize}
\item Aim to identify patterns of propagation similar to the work described in \cite{iribarren2011affinity}
\item Validate virality upon time with an easy to calculate criterion similar to metrics described in works \cite{berger2012makes,goyal2010learning,ienco2010meme,tucker2014reach,guerini2011exploring,hansen2011good}
\item Expand the concept of information diffusion across multiple OSNs, unlike the works described in \cite{yang2010predicting,kwak2010twitter,bonchi2011influence,romero2011influence}, and \cite{karnik2013diffusion} that deal with the same concept but within one OSN
\end{itemize}

\section{Methodology and Dataset}

For our study, we need domains that provide viewership counters in order to monitor and examine information propagation features. ImgUr and YouTube count views according to the user’s IP address, Reddit only provides the voting count of a post. As such, the sum of negative and positive votes is used as the Reddit views counter. We were aware that both the absence of an IP based view counter and the method Reddit uses to fuse voting, introduce some inaccuracy in our results. However, since our main aim is toward propagation rather than viewership volume, such kind of inaccuracy does not affect our initial perception of information flows.

In some preliminary tests we performed on Facebook and Google Plus, we observed that both demonstrated a fairly low propagation even on viral content. This is because both APIs of the above mentioned OSNs do not provide access to private posts. Thus, only public posts were taken into consideration. Additionally, the network usage of Google Plus is fairly low compared to Facebook. After considering these factors, we decided to include only Facebook in our analysis. Finally, in order to present our results in a consistent format between different OSNs, we address all kinds of viewership count as "Units of Interest" (UoI). A single Unit of Interest is equal to one content view in its domain (ImgUr or YouTube), one vote in Reddit, one mention in Twitter or one mention/like in Facebook.

To discard content that is not gaining attention rapidly, we employed a simple rule (mentioned hereafter as check criterion). In this rule, we examined if "Units of Interest" in Reddit are doubled in absolute numerical values for every check, for the first 4 checks after the creation of the post, as required by our virality criterion. Ou proposed criterion can be applied in real time with only a few calculations, a key difference from existing methods . Only only a small percentage of all posts reached high viewership counts within such a short time span.  0.66\% of the posts analyzed actually passed this UoI check criterion in our dataset (682 out of nearly 102400 posts).

\section{Results}

In this section, we present and analyse the results derived from our scraping procedure. Data is separated based on the the hosting domain. Reddit posts include topics such as "AdviceAnimals", "Aww", "Eathpon", "Funny", "Gaming", "Gif", "Movies", "Music", "Pics", "TIL", "Videos" and "WTF", and all were discovered with our scraping period.

\subsection{Posts and Counters}
We scrapped two different subreddits, "new" and "rising". In total, we found nearly 45000 ImgUr and YouTube posts in the "rising" category and nearly 70000 in the "new" category. More specifically, the posts of the "rising" category revealed 40966 links to ImgUr and 3084 links to YouTube. Similarly, the distribution for posts from the "new" category for ImgUr and YouTube was 51984 and 6366 respectively.

We also observed that the "new" subreddit content is more likely to become viral compared to the content in "rising" subreddit. Content featured in "new" subreddit was twice as likely to pass our UoI check, while content in "rising" category presented a higher initial UoI count but fell short on our criterion (which is doubling the UoI count in the first four check). Out of 40966 posts linking to ImgUr in "rising" subreddit, only 200 satisfied our check criterion (0.48\%). As for posts linking to YouTube, out of 3084 posts in total, only 14 satisfy our check (0.45\%). In contrast, in "new" subreddit, out of the 51984 posts that linked to ImgUr, 431 doubled their UoI count in 4 subsequent checks (0.82\%). Similarly, the number of posts that linked to YouTube and successfully passed our UoI criterion, was 37 out of 6366 (0.58\%).

The UoI criterion presented is not a virality validation, yet it provides strong evidence that the posts analysed have been viewed by enough viewers, within a small time-frame and the interest is not diminishing. Among 682 highly viewed posts, 631 were linked to ImgUr and 51 were linked to YouTube.

As mentioned in \cite{allsop2007word,berger2012makes}, positive and entertaining content was found as the most frequently shared topic. In our case, we could label "AdviceAnimals", "Funny" and "WTF topics", as entertaining content. "Aww" and "Eathpon" can be considered as containing emotive content, as described in \cite{porter2006subservient}. "TIL" is mainly informative, while "Pics" and "Videos" characterize mixed content types. "Gaming", "Movies" and "Music" contain entertainment content.

Entertaining content and positive content are shared more frequently than anything else. Although provocative content and controversial content can be created within Reddit and ImgUr, they rarely get enough votes to appear on the front page. Most of the time, they appear in the form of a "news" or "TIL" post. The number of "gaming" posts that passed our check criterion is descriptive of the interests of Reddit users, especially when compared to other forms of established entertainment such as movies and music. Out of the 682 posts, 371 were mentioned in Twitter, 7 in Facebook and 172 in both OSNs.

Posts of "funny" topic are the most shared in both Twitter and Facebook, but are not included among the top-shared topics. Additionally, almost every category presents high Twitter and low Facebook sharing percentage (a direct result of the Facebook API limitations). "Eathpon" posts with photographs from around the world were 100\% shared in Twitter. The majority of "Music" and "TIL" posts are shared in both OSNs, but the number of posts in these categories is low. In contrast, 50\% of of viral posts from "Videos" and "Movies" are shared on all OSNs, confirming the positive attitude of Internet users towards multimedia content. 

Topics with high number of viral posts ("AdviceAnimals", "Aww", "Funny", "Gaming" and "Pics"), were found to have the highest sharing percentage for both OSNs pertaining to "Pics" post from "rising" subreddit (41.38\%), followed by "Funny" posts from "new" subreddit (29.75\%). In contrast, "Aww" and "Gaming" posts from the "new" subreddit had the lowest percentages (10.53\% and 13.04\% respectively). Posts from "new" subreddit were more often shared in Twitter compared to posts from "rising" subreddit, for every topic except "Funny".

\subsection{Online Social Network Propagation}

UoI checks were simultaneously conducted on (at least) an hourly-basis for Reddit, Twitter, Facebook and the domain the image or video was hosted (ImgUr or YouTube). Thus, we were able to observe the propagation of a certain Reddit post and content in both OSNs. We wanted to measure a slow-rate information propagation from its source to both OSNs. In fact, such a propagation pattern was indeed found and was the most common propagation standard throughout our analysis of the posts and topics.

\begin{figure}[h]
\includegraphics[width=1\textwidth]{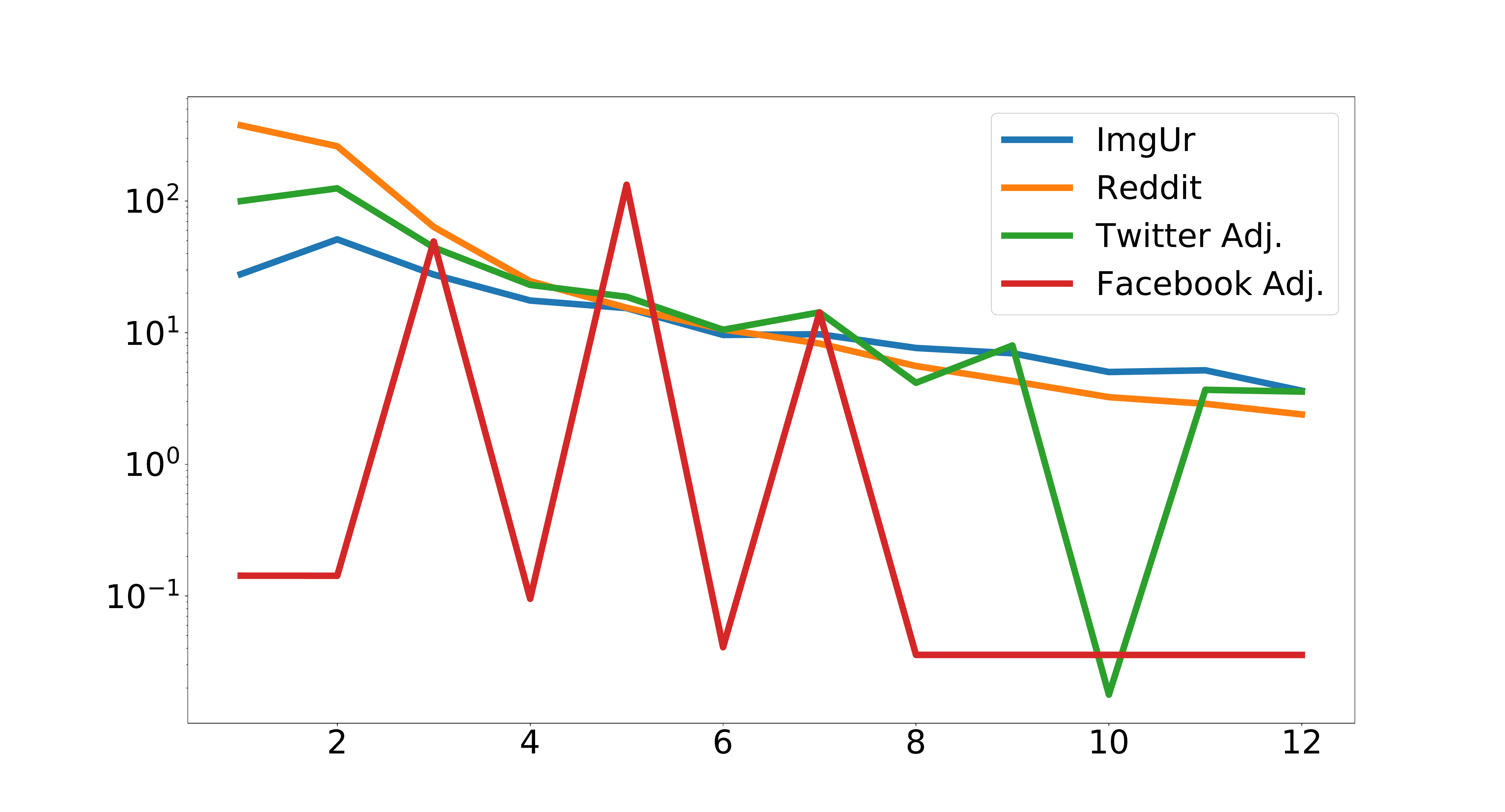}
\caption{Percent UoI Variation over time, ImgUr}
\label{fig:capture}
\end{figure}

The time interval between subsequent checks performed in the tested domains (Reddit, ImgUr or YouTube, Twitter and Facebook) is at least one hour and the first check of every post has zero UoI. Each post was monitored for 7 days. We found that after 14 checks, the interest (in terms of UoI variation) declines significantly. The mean number of checks was 65 per post, while the 14th check happened within a 24-hour period from the creation of the post.

All posts are filtered to include those that manage to double their UoI (in absolute numerical values) on Reddit on an hourly basis for a 4-hour time-frame. Consequently, every post we analysed had high Reddit UoI variance for the first 5 checks, but only a few of them had higher levels of variance after that point. As far as our OSN data scraping is concerned, we only had access to information from the public API of Facebook, while on Twitter we could only check messages within a timespan of a week due to Twitter API limitations.

Figure 1 presents the UoI variation over time for ImgUr, where its UoI are growing at a high rate concurrently with Reddit, followed by Twitter with a relatively small level of interest, and Facebook with delayed and low interest. Similar patterns appeared in all of the distinct categories, while in most of the cases Facebook had nearly zero level of interest. This specific pattern was the most common in our research and appeared in 14 out of 18 ImgUr topics. Topics such as "AdviceAnimals", "Aww, "Funny", "Eathpon", "Gif" "Pics" and "WTF", followed this pattern - in both "rising" and "new" subreddits. However in both "Gif" categories, interest variation in Reddit is slightly higher than Domain.

\begin{figure}[h]
\includegraphics[width=1\textwidth]{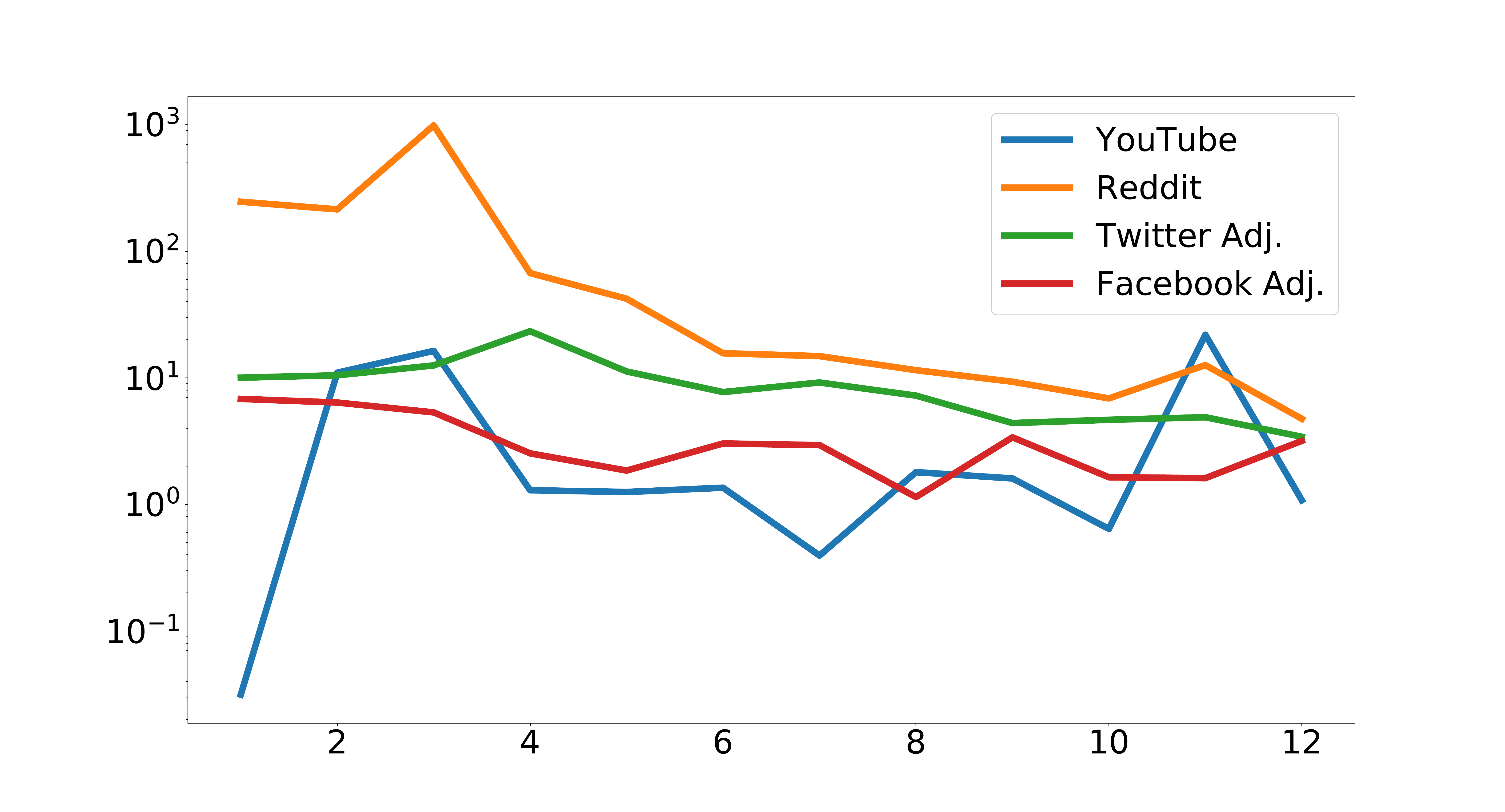}
\caption{Percent UoI Variation over time, YouTube}
\label{fig:capture}
\end{figure}

On YouTube content, we encountered a completely different propagation pattern, Figure 2. In most cases, domain UoI was not seriously affected by the increased interest in OSNs and Reddit. This is mainly because shared content in YouTube is not necessarily new, but a video that existed for a long time and becomes mildly popular over time. In our evaluation, the videos under study presented a very high initial UoI count which resulted in a low UoI variance. Apart from that, we observed high UoI variance in Reddit and Twitter, low UoI variance in Facebook, and practically zero UoI variance in the YouTube.

\section{Discussion - Future Work}

Our analysis verified many observations of previous researches mainly with respect to the micro and macro effects of information flows \cite{lazarsfeld1948peoples,katz1966personal,karnik2013diffusion}. More specifically, a single post (micro effect) in the parent domain connected with a post in Reddit, starts to accumulate views (macro effect) up to a point where the information hops to OSNs (first in Twitter and then in Facebook) flowing through individuals and eventually the interest dies off. Positive content is confirmed to be the mostly shared content in OSNs as mentioned in \cite{porter2006subservient,berger2012makes}. Similarly to \cite{ienco2010meme}, we found out that a Reddit post is most popular within the first few hours after its creation. This effect is also enhanced by the classification method of Reddit, where new posts need fewer votes (compared to old posts) to move to the first page.

As mentioned, we have observed that the most shared posts are positive and entertaining content as observed in \cite{allsop2007word,berger2012makes}. These categories include post from topics such as "AdviceAnimals", "Funny" and "WTF topics, while we also identified emotive posts, as described in \cite{porter2006subservient} on topics such as "Aww" and "Eathpon". Unfortunately, not enough viral posts of such content were found, in order to allow the identification of their propagation model. "Gaming", "Movies" and "Music" contain user-centric and very specific entertainment content, with low post appearance but a high viral to total posts ratio, while, "TIL" is mainly informative and is rarely used in posts linking to either ImgUr or Youtube.

The fact that only public Facebook can be scraped is a big factor to those particular UoI. This means that content posts and view counts on Facebook are significantly higher. Nonetheless, the approximate propagation timeframe should not be affected because of the ripple effect of social interactions \cite{long2003development}. To summarize, the most notable results of our research are:

\begin{itemize}
\item Online Social Networks are connected via shared information
\item Shared content is mostly positive and positive content is more viral
\item Shared (user generated) content has a small lifetime
\item Low percentage of content can be considered as viral
\item Information flows have micro and macro effects across networks they traverse
\item API limitations prevent further diffusion analysis
\end{itemize}

Information starts within a domain and hops to various other media within minutes. However, upon its propagation, information flows within the original domain do not stop, but slow down within a small period of time. Entertainment content and positively emotive content are the most "viral", and persistence was only found in gaming posts. Movies related posts were the only ones that spread nearly simultaneously to every domain and OSN. We should consider a new perception of information flow, one that would accurately describe content sharing in modern OSNs.

The research community can examine what kinds of social media platforms play the role of information broker for other ones, as well as what kinds of information their users prefer to link across different media and why. Are some social media not capable enough of making a post viral? Is the linkage made intentionally, or users are biased? On the other hand, social media managers can investigate probable common information channels across other social networking services, reveal groups of users with similar interests, or even compare usage and market share (market position). Finally, one can also identify the distinct role of any OSN with respect to information propagation and information lifespan (e.g. in contrast to Twitter users, users of Reddit use the service for not time-sensitive posts like bookmarking), by utilizing similar research methods.

In our next step, we will consider precise diffusion analysis relative to time. Since we have already confirmed the connection of several OSNs in this work, in the future we will look into the time frames of each OSN diffusion. But such analysis would require data timestamps and an algorithm designed for speed. Unfortunately our serial algorithm was designed without any time considerations and lacked the ability to simultaneously analyse thousands of posts. We plan to refine our scraping and analysis  method to achieve those goals. Our long-term objective is to create a viral post tool that can analyse every OSN available. We hope that future data from these OSNs will not be locked behind pay-walls and data access would be unobstructed, so that every scientist will be able to analyse these unique large-scale, content-based social networks.

Additionally, we also plan to analyse a wider range of topics (such as news or similar, which lacked required counters to observe its diffusion) in as many OSNs as we can. We were surprised to observe that various posts in Reddit of News topic attracted users' attention within Reddit long before being reported in dedicated news sites. Similarly, more sites with unrestricted APIs could be added to our analysis tool so that we can analyse similar information diffusion metrics. Diversity of topics and sites does not affect virality analysis but it would however have an impact on the complexity of our algorithm, while being able to provide more interesting results. In the future, we will investigate the aforementioned issues with a strong focus on virality, information flows and time.

\newpage

\bibliographystyle{unsrt}
\bibliography{paper} 

\end{document}